\documentclass[a4paper]{jpconf}
\usepackage{amssymb}
\usepackage{amsthm}
\usepackage{latexsym}
\newcommand{\be}{\begin{eqnarray}}
\newcommand{\ee}{\end{eqnarray}}
\newcommand{\bdm}{\begin{displaymath}}
\newcommand{\edm}{\end{displaymath}}
\newtheorem{thm}{Proposition}

\def\d{{\mathrm{d}}}

\def\generic{{\odot}}
\def\Rtt{{ R_{\hat t\hat t} }}
\def\Gtt{{ G_{\hat t\hat t} }}

\begin{document}
\title{\textbf{Cosmological milestones and energy conditions}}
\author{\textbf{C\'eline Catto\"en and Matt Visser}}
\date{}
\address{School of Mathematics, Statistics, and Computer Science, \\
Victoria University of Wellington, \\
P.O.Box 600, Wellington, New Zealand}
\ead{\textbf{celine.cattoen@mcs.vuw.ac.nz, matt.visser@mcs.vuw.ac.nz}}
 \vspace{1cm}
\begin{abstract}
Until recently, the physically relevant singularities occurring in FRW cosmologies
had traditionally been thought to be limited to the Òbig bangÓ, and possibly a
Òbig crunchÓ. However, over the last few years, the zoo of cosmological singularities
considered in the literature has become considerably more extensive, with Òbig ripsÓ
and Òsudden singularitiesÓ added to the mix, as well as renewed interest in non-singular
cosmological events such as ÒbouncesÓ and ÒturnaroundsÓ. In this talk, we present
an extensive catalogue of such cosmological milestones, both at the kinematical and dynamical
level. First, using generalized power series, purely kinematical definitions of these cosmological
events are provided in terms of the behaviour of the scale factor $a(t)$. The notion of a
Òscale-factor singularityÓ is defined, and its relation to curvature singularities (polynomial
and differential) is explored. Second, dynamical information is extracted by using the Friedmann
equations (without assuming even the existence of any equation of state) to place constraints
on whether or not the classical energy conditions are satisfied at the cosmological milestones. 
Since the classification is extremely general, and modulo certain technical assumptions complete,
the corresponding results are to a high degree model-independent.
 \end{abstract}

\section{Introduction}
The physically relevant singularities in cosmology have traditionally been thought to be limited to the big bang or big crunch~\cite{MTW}. However, over the last few years, more singularities have been considered such as big rips, sudden singularities, and there has even been a renewed interest in non-singular events such as bounces and turnarounds~\cite{rip, rip2, sudden2, sudden3, Lake}.

From observational cosmology we extract the cosmological principle, which states that the universe is homogeneous and isotropic. Hence, we consider a cosmological spacetime of the FRW form:
\begin{equation}
\d s^{2}=-\d t^{2}+a(t)^{2}\left\{\frac{\d r^{2}}{1-kr^{2}}+r^{2}\;[\d\theta^{2}+\sin^2\theta\; \d\phi^{2}]\right\},
\end{equation}
where $a(t)$ is the scale factor of the universe and $k=\pm 1, 0$ is the space curvature.

 Assuming the applicability of the Einstein equations of general relativity, we can then deduce the Friedmann equations. 
In geometric units where 
$8\pi G_N=1$ and $c=1$, we have
\begin{equation}
 \rho(t)=3\left(\frac{\dot a^{2}}{a^{2}}+\frac{k}{a^{2}}\right),
\label{Friedmann1}
\end{equation}
\begin{equation}
p(t)=-2\,{\ddot a\over a}-{\dot a^2\over a^2} - {k\over a^2},
\label{Friedmann2}
\end{equation}
\begin{equation}
\rho(t)+3\,p(t)=-6\,{\ddot a\over a},
\label{Friedmann3}
\end{equation}
and the related conservation equation
\begin{equation}
 \dot\rho(t)\;a^{3}+3\,[\rho(t)+p(t)]\,a^{2}\,\dot a=0,
\label{conservation}
\end{equation}
where any two of equations (\ref{Friedmann1}, \ref{Friedmann2}, \ref{Friedmann3}) imply (\ref{conservation}).

In this talk we shall derive generic definitions of all the singularities considered above (which we shall refer to as cosmological milestones) and we shall pin down necessary and sufficient conditions for their occurence both at the level of kinematics and dynamics.

\begin{itemize}
\item For the kinematical approach, we develop precise definitions of these cosmological milestones in terms of the scale factor $a(t)$, and with the notion of scale factor singularity in hand we check whether these milestones are true curvature singularities.
\item For the dynamical approach, we use the Friedmann equations to analyze the classical energy conditions in the vicinity of these milestones in a model-independent manner.
 \end{itemize}

\section{Kinematical analysis}
\subsection{Cosmological milestones definitions} 
From a mathematical point of view, solutions of a problem can often be expanded in Taylor series or Laurent series around their singular points. Hence, the general idea is to expand the scale factor $a(t)$ in generalized power series, similar to a Puisieux series, in the vicinity of the cosmological milestones.

\paragraph{Generic cosmological milestone:} 
Suppose we have some unspecified generic cosmological milestone,  that  is defined in terms of the behaviour of the scale factor $a(t)$, and which occurs at some finite time $t_\generic$. We will assume that in the vicinity of the milestone the scale factor has a (possibly one-sided) generalized power series expansion of the form
\begin{equation}
\label{E:power}
a(t) = c_0 |t-t_\generic|^{\eta_0} + c_1  |t-t_\generic|^{\eta_1} + c_2  |t-t_\generic|^{\eta_2} 
+ c_3 |t-t_\generic|^{\eta_3} +\dots
\end{equation}
where the indicial exponents $\eta_i$ are generically real (and are often non-integer) and without loss of generality are ordered in such a way that they satisfy
\begin{equation}
\eta_0<\eta_1<\eta_2<\eta_3\dots
\end{equation}
Finally we can also without loss of generality set
\begin{equation}
c_0 > 0.
\end{equation}
There are no \emph{a priori} constraints on the signs of the other $c_i$, though by definition $c_i\neq0$.

The first term of the right hand side of equation (\ref{E:power}) is the dominant term, and is therefore responsible for the convergence or divergence of the scale factor at the time $t_\generic$. The indices $\eta_i$ are used to classify the cosmological milestones and the absolute values are used to distinguish a past event from a future event. This power series expansion of the scale factor is sufficient to represent almost all the physical models that we are aware of in the literature. Table \ref{table1} represents this cosmological milestone classification depending on the value of the scale factor.
\begin{table}[htdp]
\caption{Classification of cosmological milestones}
\begin{center}
\begin{tabular}{||c||c||c||}    \hline
Cosmological & Scale factor    & Indices  \\
milestones &  value                    & $\eta_i$ \\ \hline \hline
Big Bang/  &  $a(t_\generic)=0$   &  $\eta_0>0$ \\
Big Crunch&                                 &                  \\
 \hline
Sudden       & $a(t_\generic)=c_0$ & $\eta_0=0$ \\
Singularity   &$a^{(n)}(t_\generic)=\infty$& $\eta_1$ non-integer \\   \hline 
Extremality & $a(t_\generic)=c_0$  &  $\eta_0=0$ \\
events         &                                  & $\eta_i \in  \mathrm{Z^+}$ \\   \hline
Big rip           &$a(t_\generic)=\infty$ & $\eta_0<0$\\ \hline 
\end{tabular}
\end{center}
\label{table1}
\end{table}
 For Big bangs or Big crunches the scale factor is zero and it corresponds to $\eta_0>0$ in terms of the power series expansion indices. The value $\eta_0=0$ leads to the scale factor being finite and describes two types of milestones, sudden singularities and extremality events --- which are not singularities in any sense (bounce, turnaround, inflexion). Specifically, sudden singularities are of order $n$ where the $n^{th}$ derivative of  the scale factor is infinite:
\begin{equation}
a^{(n)}(t\to t_\generic) \sim c_0 \; \eta_1 (\eta_1-1)(\eta_1-2)\dots (\eta_1-n+1) \; |t-t_\generic|^{\eta_1-n}\to\infty,
\end{equation}
and therefore $\eta_1$ has to be a non-integer~\cite{sudden2, sudden3}.
In contrast for extremality events, the power series expansion of the scale factor $a(t)$ is a simple Taylor series with $\eta_i \in \mathrm{Z^+}$. Furthermore, an extremality event is of order $n$ where:
\begin{itemize}
\item $a^{(2n)}(t_\generic)>0$ for a bounce (local minimum)~\cite{bounce, Tolman},
\item $a^{(2n)}(t_\generic)<0$ for a turnaround (local maximum),
\end{itemize}
so that for a bounce or turnaround we write:
\begin{equation}
a(t) = a(t_\generic) + {1\over(2n)!} a^{(2n)}(t_\generic)\; [t-t_\generic]^{2n} + \dots;
\end{equation}
\begin{itemize}
\item $a^{(2n+1)}(t_\generic)\neq 0$ for an inflexion event (neither local minimum nor maximum),
\end{itemize}
so that:
\begin{equation}
a(t) = a(t_\generic) + {1\over(2n+1)!} a^{(2n+1)}(t_\generic)\; [t-t_\generic]^{2n+1} + \dots.
\end{equation}
Finally, a big rip occurs when the scale factor $a(t)$ is infinite at some finite time $t_\generic$~\cite{rip, rip2}, it corresponds to $\eta_0<0$ in terms of the power series expansion. Even though we distinguish a ``future rip'' from a ``past rip'', the literature to date has only considered future rips as a past rip would be a very unusual beginning to the universe.

\paragraph{Summary:} We have written a generic definition of the scale factor $a(t)$ based on generalized power series (generalized Puisieux series) for all the physically relevant cosmological milestones found in the literature to date (big bang, big crunch, sudden singularity, extremality events and big rip). We can now use the parameters to explore the kinematical and dynamical properties of the cosmological milestones, see whether they are true curvature singularities and whether the energy conditions hold in the vicinity of the time of the event $t_\generic$. Note that for most calculations it is sufficient to use the first three terms of the power series expansion. 

\subsection{Spacetime curvature}  \label{spacetime_curvature}
To determine whether a cosmological milestone is a true spacetime curvature singularity or not, in the sense of classical general relativity, we need to have a closer look at the Riemann tensor at the time of the event.
However, due to the symmetries of the FRW geometry we are assuming, the Weyl tensor vanishes and hence it is sufficient to look at the Ricci tensor. Now if we take into account the spherical symmetry and translational symmetry, we only need to look at the only two non-zero orthonormal independent components of the Ricci tensor to test for curvature singularities:
\begin{equation}
R_{\hat t\hat t} \qquad\hbox{and} \qquad R_{\hat r\hat r}=R_{\hat\theta\hat\theta}=R_{\hat\phi\hat\phi}.
\end{equation}
On another hand, one could consider instead $R_{\hat t\hat t}$ and the Ricci scalar $R$, or $R_{\hat t\hat t}$ and the Einstein $\hat t\hat t$-component $\Gtt$, or alternatively the Ricci scalar $R$ and the combination $R_{ab}R^{ab}$.

\paragraph{Polynomial curvature singularity:}
\begin{thm}
 To characterize all polynomial curvature singularities in a FRW spacetime geometry, and  hence decide whether the cosmological milestones we have defined are true curvature singularities, it is sufficient to test $R_{\hat t\hat t}$ and $\Gtt$ in orthonormal components for finiteness:
\begin{equation}
\Rtt = - 3 \;{\ddot a\over a}; \qquad
\Gtt = 3 \left( {\dot a^2\over a^2} + {k\over a^2}\right).
\end{equation}
\end{thm}
Note that $\Rtt$ is independent of the space curvature $k$,  $\Gtt$ is independent of the second derivative of the scale factor $\ddot{a}$, and finally $\Rtt$ and $\Gtt$ are linearly independent.

From the definition of a generic cosmological milestone (assuming $t>t_\generic$ for simplicity) we have:
\begin{eqnarray}
a(t) &=& c_0 (t-t_\generic)^{\eta_0} + c_1  (t-t_\generic)^{\eta_1} +\dots   \label{def_a} \\
\dot a(t) &=& c_0 \eta_0 (t-t_\generic)^{\eta_0-1} + c_1  \eta_1 (t-t_\generic)^{\eta_1-1}  +\dots  \label{def_adot} \\
\ddot a(t) &=& c_0 \eta_0 (\eta_0-1) (t-t_\generic)^{\eta_0-2} 
+ c_1  \eta_1 (\eta_1-1) (t-t_\generic)^{\eta_1-2}  \nonumber \\
&&+ c_2  \eta_2  (\eta_2-1) (t-t_\generic)^{\eta_2-2}  
+\dots \label{def_addot} 
\end{eqnarray}
The method we use to study the finiteness of $\Rtt$ or $\Gtt$ is generic. We write $\Rtt$ in the vicinity of a cosmological milestone using the power series expansion of the scale factor and its derivatives, keeping only the most dominant terms:
\begin{itemize}
\item if $\eta_0\neq0$ and $\eta_1\neq1$ then
\begin{equation}
\Rtt = - 3 \;{\ddot a\over a}  \sim -3 \; {\eta_0(\eta_0-1)\over (t-t_\generic)^2}, \quad
\hbox{hence} \quad \lim_{t\to t_\generic} \Rtt = \mathrm{sign}(\eta_0[1-\eta_0])\, \infty
\end{equation}
\item if $\eta_0=0$ and $\eta_1\neq1$ then
\begin{equation}
\Rtt   \sim  -3 \; {\eta_1(\eta_1-1)c_1\over c_0} \;  (t-t_\generic)^{\eta_1-2}, 
\quad \hbox{hence} \quad
\lim_{t\to t_\generic} \Rtt = \left\{ \begin{array}{ll}
0 &  \quad \eta_1 > 2;\\
-6c_1/c_0 &  \quad \eta_1 = 2;\\
\mathrm{sign}(c_1)\infty &  \quad 
\eta_1\in(1,2);\\
-\mathrm{sign}(c_1)\infty &\quad  \eta_1 \in(0,1);\\
\end{array}
\right.
\end{equation}
\item if both $\eta_0=0$ and $\eta_1= 1$ then
\begin{equation}
\Rtt   \sim  -3 \; {\eta_2(\eta_2-1)c_2\over c_0} \;  (t-t_\generic)^{\eta_2-2},
\quad \hbox{hence} \quad
\lim_{t\to t_\generic} \Rtt = \left\{ \begin{array}{ll}
0 & \quad\eta_2 > 2;\\
-6c_2/c_0&  \quad\eta_2=2; \\
-\mathrm{sign}(c_2)\infty & \quad\eta_2\in(1,2); \\
\end{array}
\right.
\end{equation}
\item if $\eta_0=1$, then (since $\eta_1>1$)
\begin{equation}
\Rtt   \sim  -3\; {\eta_1(\eta_1-1)c_1\over c_0} \;  (t-t_\generic)^{\eta_1-3},
\quad \hbox{hence} \quad
\lim_{t\to t_\generic} \Rtt = \left\{ \begin{array}{ll}
0 &  \quad \eta_1>3;\\
-18c_1/c_0& \quad\eta_1=3; \\
-\mathrm{sign}(c_1)\infty & \quad\eta_1\in(1,3); \\
\end{array}
\right.
\end{equation}
\end{itemize}
Now we can see that the number of cases in which $\Rtt$ remains finite is rather limited. Explicitly, $\Rtt$ is finite provided that:
\begin{itemize}
\item $\eta_0=0$, $\eta_1\geq 2$;
\item $\eta_0=0$, $\eta_1=1$, $\eta_2\geq2$;
\item $\eta_0=1$ and $\eta_1\geq 3$.
\end{itemize}
We apply the same generic argument to analyze $\Gtt$ in the vicinity of a cosmological milestone. We have:
\begin{itemize}
\item if $ \eta_0\neq0$ then
\begin{equation}
\Gtt \sim 3 \left[ {\eta_0^2\over(t-t_\generic)^2} + {k\over c_0^2(t-t_\generic)^{2\eta_0}} \right],
\end{equation}
\item while if $\eta_0=0$ then
\begin{equation}
\Gtt \sim 3 \left[ {\eta_1^2c_1^2(t-t_\generic)^{2(\eta_1-1)} + k\over c_0^2} \right].
\end{equation}
\end{itemize}
After a few technical calculations we find that the necessary and sufficient conditions for $\Gtt$ to remain finite are:
\begin{itemize}
\item $\eta_0=0$, $\eta_1\geq 1$;
\item $\eta_0=1$, $c_0^2+k=0$, and  $\eta_1\geq 3$.  \\
(Note that $c_0^2+k=0$ implies $k=-1$, and $c_0=1$.)
\end{itemize}
To decide the characterization of a cosmological milestone as a polynomial curvature singularity, we need to combine the two previous results.
\begin{thm}
Assuming we can write the scale factor as a generalized power series in a the vicinity of a cosmological milestone, the necessary and sufficient conditions for both $\Gtt$ and $\Rtt$ to remain finite, so that this cosmological milestone is \emph{not} a polynomial curvature singularity, are:
\begin{itemize}
\item $\eta_0=0$, $\eta_1\geq 2$;
\item $\eta_0=0$, $\eta_1=1$, $\eta_2\geq2$;
\item $\eta_0=1$, $k=-1$, $c_0=1$, and  $\eta_1\geq 3$.
\end{itemize}
\end{thm}
Note that the first two cases concern a sub-class of sudden singularities ($\eta_0=0$) and the last case corresponds to a type of big bang or big crunch ($\eta_0=1$). This result seems quite surprising, however, one can ask whether those exceptional cases are excluded if we generalize the idea of polynomial curvature singularity to that of \emph{derivative} curvature singularity.

\paragraph{Derivative curvature singularity:}
Even if the Riemann tensor does not blow up in the vicinity of a cosmological milestone, does this property still hold for some derivative of the Riemann tensor?
We define a $n^{th}$-order derivative curvature singularity by the condition that the $n^{th}$ derivative of the curvature tensor blows up.
Because of the symetries of the FRW universe that we mentioned earlier the only derivatives of interest are time derivatives, and hence we only need to consider:
\begin{eqnarray}
{\d^n \Rtt\over \d^n t}& =& - 3 {a^{(n+2)}\over a} + \hbox{ (lower-order derivatives)};\\
 {\d^n \Gtt\over \d^n t} &=& 3 {\dot a \; a^{(n+1)}\over a^2}  + \hbox{ (lower-order derivatives)}.
\end{eqnarray}
\begin{thm}
To avoid a $n^{th}$-order derivative curvature singularity in the vicinity of a cosmological milestone, we must force all ${a^{(j)} \over a}$ with $j \leqslant n+2$ to remain finite. Subsequently, the indices $\eta_i$ in the generalized power series expansion of the scale factor $a(t)$ must be positive integers forcing the series to be a Taylor series.
\end{thm}

\paragraph{Conclusion: Are cosmological milestones true curvature singularities?}
Assuming we can write $a(t)$ as a generalized power series in the vicinity of some cosmological milestone, we have demonstrated that, with the exception of two very specific cases, almost all these events are true curvature singularities in the sense that they are both polynomial and derivative curvature singularities. The two situations in which a cosmological milestone is \emph{not} a true curvature singularity are if:
\begin{itemize}
\item $\eta_0=0$,\; $\eta_{i}\in Z^+$; corresponding to an extremality event (bounce, turnaround, or inflexion event) rather than a bang, crunch, rip, or sudden singularity;
\item $\eta_0=1$, $k=-1$, $c_0=1$, $\eta_{i}\in Z^+$, and  $\eta_1\geq 3$; corresponding to a FRW geometry that smoothly asymptotes near the cosmological milestone to the Riemann-flat  Milne universe\footnote{The case $a(t)=t$, $k=-1$ is the Milne universe and is a disguised portion of Minkowski space.} which is not a really popular model in cosmology.
\end{itemize}
\section{Dynamical analysis}
We now wish to explore the consequences of the previous kinematical results at a dynamical level using the Friedmann equations (\ref{Friedmann1}, \ref{Friedmann2}, \ref{Friedmann3}) and (\ref{conservation}). How ``strange'' does physics get in the vicinity of some cosmological milestone? To answer this question we ask what happens to the energy conditions of general relativity at the cosmological milestones. Even though one can question if they are truly fundamental~\cite{twilight}, they  help quantifying how ``strange'' physics gets at the cosmological events by informing us as to which type of physical conditions are satisfied or not.

The energy conditions of general relativity are the \emph{null}, \emph{weak}, \emph{strong}, and \emph{dominant} energy conditions. Assuming a FRW universe and making use of the Friedmann equations (\ref{Friedmann1}, \ref{Friedmann2}, \ref{Friedmann3} and \ref{conservation}), they specialize to the following conditions~\cite{bounce, Tolman, twilight}:
 \begin{itemize}
\item{}The [NEC] is satisfied if $\rho+p \geq 0$ that is, in view of the Friedmann equations, if $k \geq a \;\ddot a - \dot a^2$.
\item{} The [WEC] is satisfied if the NEC is and if in addition $\rho\geq0$, that is $ k \geq - \dot a^2$.
\item{}The [SEC] is satisfied if the NEC is and if in addition $\rho+3p\geq0$, that is $\ddot a \leq 0$.
\item{}The [DEC] is satisfied if $\rho\pm p \geq 0$, which reduces to the NEC and the condition $k \geq -(a\; \ddot a +2\dot a^2)/2$.
\end{itemize}
Note that the NEC is the most interesting condition as it is the weakest of all and it leads to the strongest theorems (e.g. Hawking area theorem for black hole horizon). The DEC is usually satisfied for ``normal matter'' and can be interpreted as saying that the speed of energy flow of matter is always less than the speed of light.

We now consider the generalized power series expansion for the scale factor $a(t)$ defined in (\ref{def_a}) and the consecutive derivatives (\ref{def_adot}) and (\ref{def_addot}). Following the same pattern detailed in section \ref{spacetime_curvature} for $\Rtt$, we obtain a complete classification of all the energy conditions as being satisfied or not depending on the different values of the space curvature $k$, the indices $\eta_i$, and the coefficients $c_i$ (see~\cite{Cattoen:2005dx} for full details). In this article we will not present the classification in full detail, but we will mention some of the main results for the different energy conditions.
\paragraph{NEC:}
The classification obtained for the NEC is in agreement with most of the results scattered throughout the literature. For example, it is obvious that big rips always violate the NEC and therefore all the other energy conditions. Note that the NEC holds for big bangs and big crunches independently of the space curvature $k$ within the range $\eta_0 \in (0,1)$. However for a class of sufficiently ``violent'' bangs ($\eta_0>1$) and hyperbolic spatial curvature ($k=-1$) the NEC is violated. The analysis becomes more tedious for the classes of sudden singularities; many but \emph{not all} of them violate the NEC.
\paragraph{WEC:}
Remember that the condition $k>-\dot{a}^2$, in addition to the NEC, must hold in order for the WEC to be satisfied. Consequently, the WEC is equivalent to the NEC concerning bangs, crunches and rips, and the only changes arise for the class of cosmological milestone corresponding to $\eta_0=0$ (sudden singularities or extremality events). The WEC also holds for big bangs and big crunches independently of the space curvature $k$ within the range $\eta_0 \in (0,1)$. Again note that many but \emph{not all} of the sudden singularities violate the NEC.
\paragraph{SEC:}
Not only do the big rips violate the NEC, WEC, and therefore the SEC, but \emph{all} ``violent'' bangs or crunches violate the SEC (independently of $k$). However, the SEC is satisfied for big bangs and big crunches independently of the space curvature $k$ within the range $\eta_0 \in (0,1)$. Also, note that turnarounds satisfy the SEC while bounces violate the SEC. Regarding inflexion events, the SEC is violated either just before or just after (in agreement with~\cite{bounce, Tolman}).
\paragraph{DEC:}
There are several classes of cosmological milestones for which the DEC holds all the way to the singularity. In particular, the DEC is satisfied for big bangs and big crunches independently of the space curvature $k$, within the more restrained range $\eta_0 \in (1/3,1)$. More surprisingly, there are a few classes of sudden singularities \emph{satisfying} the DEC. This result seems to be in contradiction with Lake's analysis~\cite{Lake}, however the disagreement resides in the definition of just ``how sudden'' is a sudden singularity. Only some sufficiently ``gentle'' sudden singularities, those for which the first divergence occur in higher derivatives $a^{(n)}(t\to t_\generic)$ for $n\geq3$, satisfy the DEC. Indeed, for some of these events the DEC can be satisfied all the way to the singularity. This result is in agreement with Barrow and Tsagas~\cite{sudden3}, though now we have a full characterization of those situations.

\section{Discussion}

Assuming a FRW universe, and applicability of general relativity, we have explored three issues concerning singularities: 
\begin{itemize}
\item We have developed a generic classification of the various physically relevant ``cosmological milestones'' occurring in the literature in terms of a generalized power series expansion of the scale factor $a(t)$.
\item With the notion of generalized power series, we have defined (on a purely kinematical level), the idea of polynomial and derivative curvature singularities. It is therefore possible to classify all ``cosmological milestones'' as to whether they are singular or not, with the class of non-singular milestones being strictly limited (to extremality events or asymptotically Milne bangs/crunches). 
\item On dynamical grounds, we have checked the validity or otherwise of the classical energy conditions in the vicinity of the cosmological milestones. Depending on one's attitude towards the energy conditions~\cite{twilight}, one could use this catalogue as a guide towards deciding on potentially interesting scenarios to investigate. 
\end{itemize}

Finally, if in the vicinity of any cosmological milestone, the input scale factor $a(t)$ is a generalized power series, then all physical observables (e.g. $H$, $q$, the Riemann tensor, \emph{etc}.) will likewise be a generalized power series. By checking the related indicial exponents that can be calculated from the indicial exponents of the scale factor, one can determine whether or not the particular physical observable then diverges at the cosmological milestone.

\ack This research was supported by the Marsden Fund administered by the Royal Society of New Zealand.

\end{document}